\documentclass[useAMS]{mn2e}
\usepackage{lscape}
\usepackage{rotating}

\def\msun{\hbox{M$_\odot$}}

\def\cm3{\hbox{cm$^{-3}$}}

\input psfig.sty
\input epsf.sty
\usepackage{graphicx}
\usepackage{epsfig}
\title[Stellar Rotational effects on cluster CMDs]
{The Effect of Stellar Rotation on Colour-Magnitude Diagrams: On the apparent presence of multiple populations in intermediate age stellar clusters}
\author[Bastian \& de Mink] {N. Bastian$^{1}$, S.E. de Mink$^{2}$\\
$^1$ Institute of Astronomy, University of Cambridge, Madingley Road, Cambridge, CB3 0HA, UK\\
$^2$ Sterrekundig Instituut, Universiteit Utrecht, PO Box 80000, 3508 TA Utrecht, The Netherlands 
}
\date{Accepted. Received; in original form}
\pagerange{\pageref{firstpage}--\pageref{lastpage}}
\pubyear{2009}
\begin{document}
\maketitle
\label{firstpage}
\begin{abstract}

A significant number of intermediate age clusters ($1-2$~Gyr) in the Magellanic Clouds appear to have multiple stellar populations within them, derived from bi-modal or extended main sequence turn offs.  If this is interpreted as an age spread, the multiple populations are separated by a few hundred Myr, which would call into question the long held notion that clusters are  simple stellar populations. Here we show that stellar rotation in stars with masses between $1.2-1.7$~\msun\ can mimic the effect of a double or multiple population, whereas in actuality only a single population exists.  The two main causes of the spread near the turn-off are the effects of stellar rotation on the structure of the star and the inclination angle of the star relative to the observer.  Both effects change the observed effective temperature, hence colour, and flux of the star.  In order to match observations, the required rotation rates are 20-50\% of the critical rotation, which are consistent with observed rotation rates of similar mass stars in the Galaxy.  We provide scaling relations which can be applied to non-rotating isochrones in order to mimic the effects of rotation.   Finally, we note that rotation is unlikely to be the cause of the multiple stellar populations observed in old globular clusters, as low mass stars ($<1$\msun) are not expected to be rapid rotators. 

\end{abstract}

\section{Introduction}

Due to the increase in spatial resolution and photometric accuracy, colour-magnitude diagrams (CMDs) of star clusters have become more refined in recent years, which has led to the discovery of unexpected phenomena.  Arguably the most exciting recent discovery in cluster research has been that clusters often appear to contain structure in their observed CMDs which cannot be accounted for by current evolutionary isochrones for a single age/metallicity (see Piotto~2008 for a recent review).  Of particular interest has been the discovery that intermediate age (1-2~Gyr) clusters in the Magellanic Clouds often have bi- or multi-modal distributions in their main-sequence turn-off  (Mackey \& Broby Nielsen~2007;  Mackey et al.~2008; Glatt et al.~2008; Milone et al.~2009).  This spread has been interpreted as an age difference of $200-300$~Myr between multiple co-existing stellar populations, which is much longer than the age spread inferred in currently forming clusters ($<3$~Myr, e.g. Massey \& Hunter 1998).  This potentially large age spread has called into question whether intermediate age star clusters are indeed simple stellar populations (i.e. all stars have similar ages and metallicities).

Various models have been put forward to explain how a cluster can contain multiple populations with age differences of a few hundred Myr.  Three recent models are: 1) the formation of a second generation of stars from the ejecta of first generation AGB stars (D'Ercole et al.~2008, Goudfrooij et al.~2009), 2) the merging of two (or more) star clusters with large age differences ($>200$~Myr; Mackey \& Broby Nielsen~2007) and 3) star cluster - GMC collisions, from which the newly formed stars in the GMC are retained in the existing star cluster (Bekki \& Mackey~2009).  Each of these rather exotic models have problems explaining certain observations.  In the model 1 scenario, the second generation can never have as many or more stars than the first (without drastically changing the stellar initial mass function between the two generations), which is in contrast with observations (Milone et al.~2009).  Additionally, this model predicts that younger clusters (100-500~Myr) should contain dense pockets of gas that are forming new stars. The main drawback of model 2 is that collisions between clusters with large age differences are quite rare, and therefore would not be expected to affect a large fraction of clusters.  In the 3rd scenario the newly formed stars are expected to form in a disk-like structure in the centre of the cluster, which is at odds with the observations that the two populations have similar spatial distributions (unless very rapid relaxation is invoked).



Alternatively, the observed spread near the turn-off may be due to stellar evolutionary effects.  
Mackey et al~(2008) demonstrate that this spread cannot be explained by
unresolved binaries.  In this work we investigate how stellar rotation
affects the CMD of these clusters.

The reduced gravity in rotating stars results in lower luminosities
and effective temperatures (e.g. Faulkner et al. 1968, Meynet \&
Maeder 1997). Also the orientation of the star with repect to the
line of sight is important: a star deformed by fast rotation appears
brighter and hotter if viewed pole-on.  Various authors have
demonstrated that these effects can significantly alter the color and
magnitude of stars (e.g. Collins \& Smith 1985; Gray \& Garrison 1987;
Townsend, Owocki, \& Howarth~2004) and therefore the CMD of a cluster
(D'Souza et al.~1992, P{\'e}rez Hern{\'a}ndez et al.~1999). In
addition, mixing processes induced by rotation modify the composition
of the stellar envelope and effectively increase the size of the
stellar core, resulting in higher luminosities and cooler temperatures
for turn-off stars (Palacios et al. 2003).

The stars near the turn-off in the intermediate-age LMC clusters, with
masses above about 1.2\msun, are expected to be fast rotators.   Many
of their galactic counterparts, late A and early F type stars are found to rotate rapidly, both in the field (e.g. Royer et al.~2007) and in open clusters (e.g.~Boesgaard~1987; Gaige~1993). One might expect
even higher rotation rates for stars in clusters with respect to field
stars (Huang \& Gies~2006, however see Huang \& Gies~2008) and for
stars in the lower metallicity environment of the LMC (Hunter et
al.~2008). Only stars near and just below the turn-off, which have
radiative envelopes, are expected to rotate fast. Lower mass stars are
found to be slow rotators, probably due to the magnetic fields they
can generate in their convective envelopes (e.g.~Cardini \&
Cassatella~2007).

In the present work, we explore how the shape of the observed CMD of
an intermediate-age ($\sim1-2$~Gyr) cluster is affected by
rotation. We employ a simple method using non-rotating isochrones and
correction factors to account for rotational effects.  We note that we
use the standard B, V, and I filter notation, although all results
correspond to the HST filters F435W, F555W, and F814W, respectively.



\section{Method}
\label{sec:method}

To model the effects of rotation on the shape of the CMD of an
intermediate age cluster we apply a simple method starting from a
standard isochrone based on non-rotating stellar models.
We adopt a Salpeter~(1955) stellar initial mass function, and sample
50,000 stars stochastically between $0.8-5$\msun. We use the BaSTI
isochrones (Pietrinferni et al.~2004) to find the HST-ACS F555W and
F814W magnitudes (hereafter V and I, respectively) for each star, assuming an age of
 1.25~Gyr and metallicity of Z=0.008, relevant for intermediate age clusters in the LMC.

We assume that the distribution of rotation rates is described by a
Gaussian distribution with a peak at $\omega=0.4$ and a standard
deviation of 0.25, where $\omega$ is the fraction of the critical
break-up rotation rate. We do not allow extreme rotation rates by
excluding $\omega > 0.7$. The precise shape of the distribution does
not affect our conclusions and we will return to this point in
\S~\ref{sec:results}.  Below 1.2\msun\ (at LMC metallicity) stars
exhibit a convective envelope, in which they may generate a magnetic
field.  The braking torque of winds magnetically coupled to the
surface can efficiently slow down the rotation rate (Schatzman 1962;
Mestel \& Spruit~1987).  Also, stars that evolve off the main sequence
expand and slow down their rotation rate due to conservation of
angular momentum.  Therefore, we assume no rotation for stars evolved
beyond the end of the main sequence and for stars with masses smaller
than 1.2\msun\footnote{This gives a mass range of
$1.2-1.65$\msun\ where rotation is considered.  We note, however, that
the development of a convective envelope depends mainly on the $T_{\rm
eff}$ (e.g.~Talon \& Charbonnel 2004), which may be a better proxy for
investigating when magnetic braking is most efficient.}.

Below we derive how rotation changes the (apparent) luminosity and
effective temperature of a rotating star. To convert these corrections
to changes in colour we use that the dependence of the V-I colour on
effective temperature in the region of interest is well described by
the relation $\Delta(V-I) = -3.3\Delta \log T_{\rm eff}$, (Eldridge,
Mattila, \& Smartt~2007) based on the BaSeL-2.2 atmosphere library by
Westera, Lejeune \& Buser~(1999).
 
We have tested the above relation by creating black body spectra of various effective temperatures and convolving them with the filter response functions of the HST-ACS F555W and F814W
filters.
This simple derivation agrees well with
the results from detailed atmosphere models of Eldridge et al.~(2007).  A cubic polynomial of the form $\Delta(V-I) / \Delta(\log T_{\rm eff}) = a + bx +
cx^2 + dx^3$, where $x = \log (T_{\rm eff})$, provides a good fit
to the d(V-I)/d($\log T_{\rm eff}$) distribution when a=-406.350,
b=263.140, c=-57.1160 \& d=4.1519.  Carrying out the same analysis for
d(B-V)/d($\log T_{\rm eff}$) (where B is the HST ACS F435W filter)
results in a=-214.177, b=132.886, c=-27.5868 \& d=1.9145.  All fits
were carried out over the range $ 3.6 < \log (T_{\rm eff}) < 4.5$.


\paragraph*{Effect on the structure} 


To quantify the effects of rotation on the structure due to the
reduced gravity and mixing we use the 1D hydrodynamic stellar
evolution code described by Heger et al.~(2000) and Yoon et
al.~(2006), which takes into account the effects of rotation on the
structure and mixing induced by rotation.  We use the same choices for
the mixing parameters as De Mink et al.~(2009), which were calibrated
by Brott et al.~(in prep). Assuming an initial composition relevant
for the LMC, we evolve stellar models with an initial mass of
1.5\msun\ from the onset of hydrogen burning, beyond the end of the
main sequence, assuming initially uniform rotation rate $\omega$ of 0,
0.23, 0.35, 0.48, 0.63 and 0.77.
For not too extreme rotation rates ($\omega < 0.7$) the effective
temperature of these models can be described with a simple fit
\[T_{\rm eff}(\omega)/T_{\rm eff}(0) = 1 - a \omega^2,\] where $T_{\rm
eff}(0)$ is the effective temperature of the non-rotating model and
$a$ is the fit coefficient which changes from 0.17 at zero age to 0.19
at the end of the main sequence. Similarly, for the luminosity we fit
\[L(\omega)/L(0) = 1 - b\omega^2,\] where $b$ varies from 0.11 to 0.03
over the main-sequence evolution.
These relative changes are consistent with results of the Geneva group
(G. Meynet, priv. comm.) for a 4\msun\ star, the lowest in their model
grid. As we are interested in the turn-off, where stars no longer
resemble zero-age stars, we adopt the corrections for the end of the
main sequence.  We note that differences between the rotating and
non-rotating model tracks for the post-main sequence evolution are
much smaller than on the main sequence, further justifying our assumption that only the main
sequence and the turn-off are strongly affected by rotation.

\paragraph*{Effect of inclination} Depending on the inclination angle
between the rotation axis and the line of sight, a star deformed by
rotation may appear brighter and hotter if viewed pole-on.  To
estimate the importance of this effect we assume that the star is
deformed according to the Roche model. To obtain the flux emitted
towards the observer for various inclinations and rotation rates,
$\omega$, we integrate the flux of each surface element, which depends
on the local effective gravity according to von Zeipel~(1924). We take
into account that each surface element is stretched and inclined with
respect to the line of sight. Furthermore we adopt a simple
limb-darkening law (assuming a plane-parallel gray atmosphere using
the Eddington approximation). We compute the apparent effective
temperature by averaging the effective temperature of each surface
element given by the von Zeipel theorem, weighted by the flux emitted
by this element toward the observer. We emphasize that the adopted
approach is simplified, but the level of sophistication is sufficient
for the objective of this paper.

We take this effect into account by assigning a random orientation
to each star in the CMD and applying the correction computed as
described above to the luminosity and effective temperature.  The
effect of the inclination is small for moderately rotating stars
($\omega<0.5$): less than $1\%$ on the effective temperature for
most viewing angles, compared to $\sim5\%$ due to structural effects.

\section{Results}
\label{sec:results}

\begin{figure*}
\includegraphics[width=8.5cm]{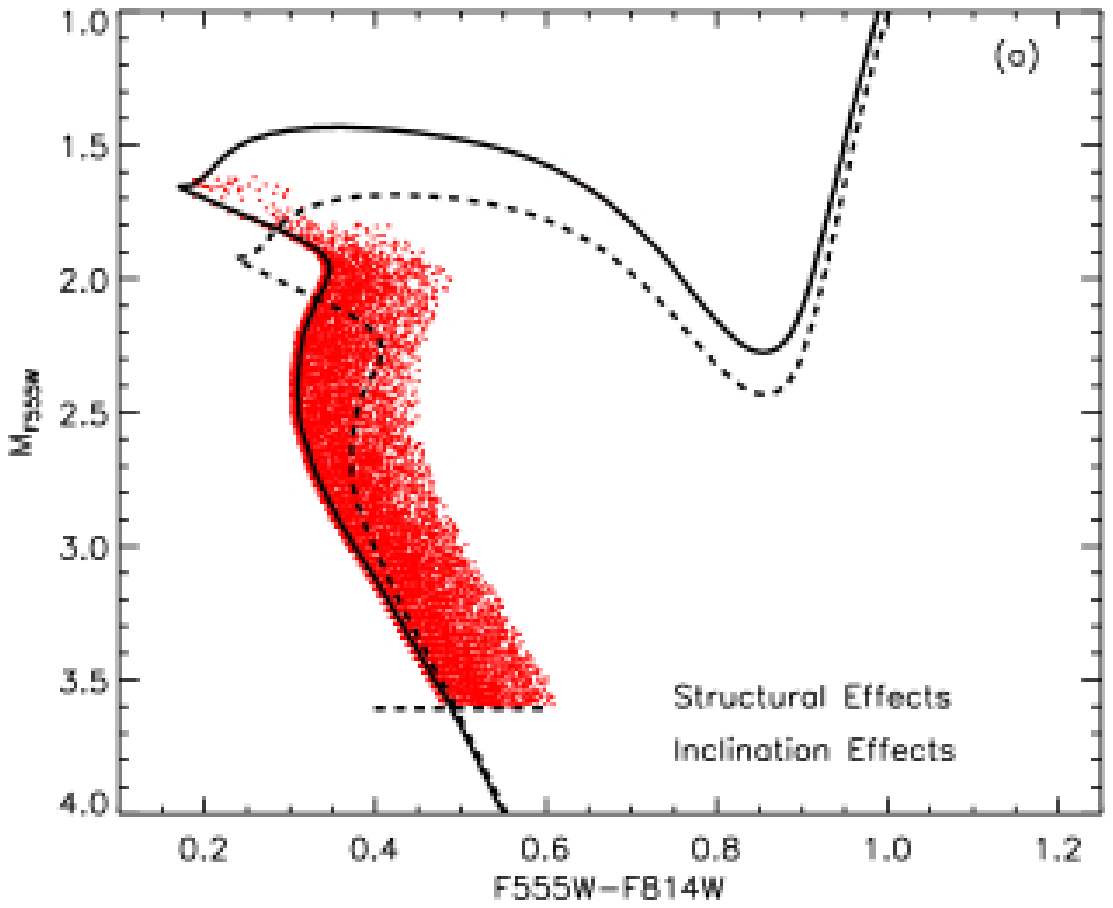}
\includegraphics[width=8.5cm]{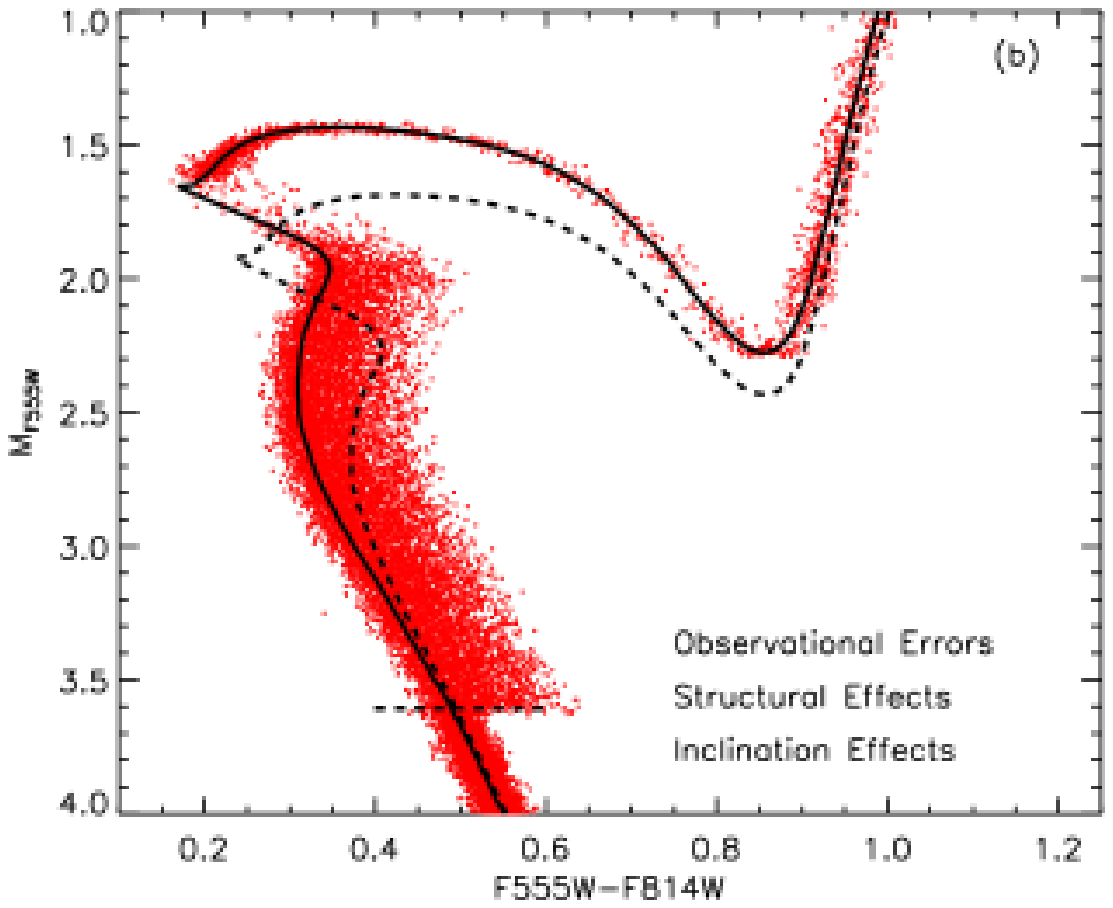}
\includegraphics[width=8.5cm]{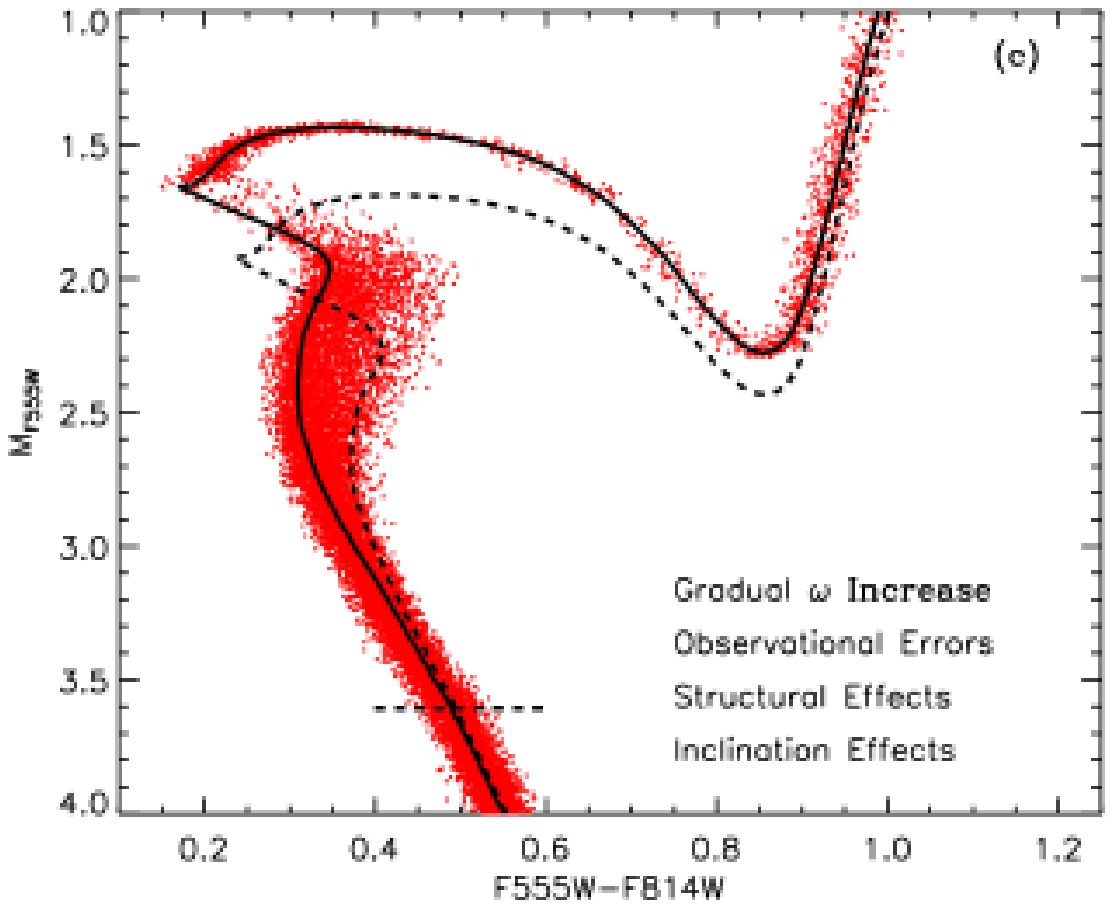}
\includegraphics[width=8.5cm]{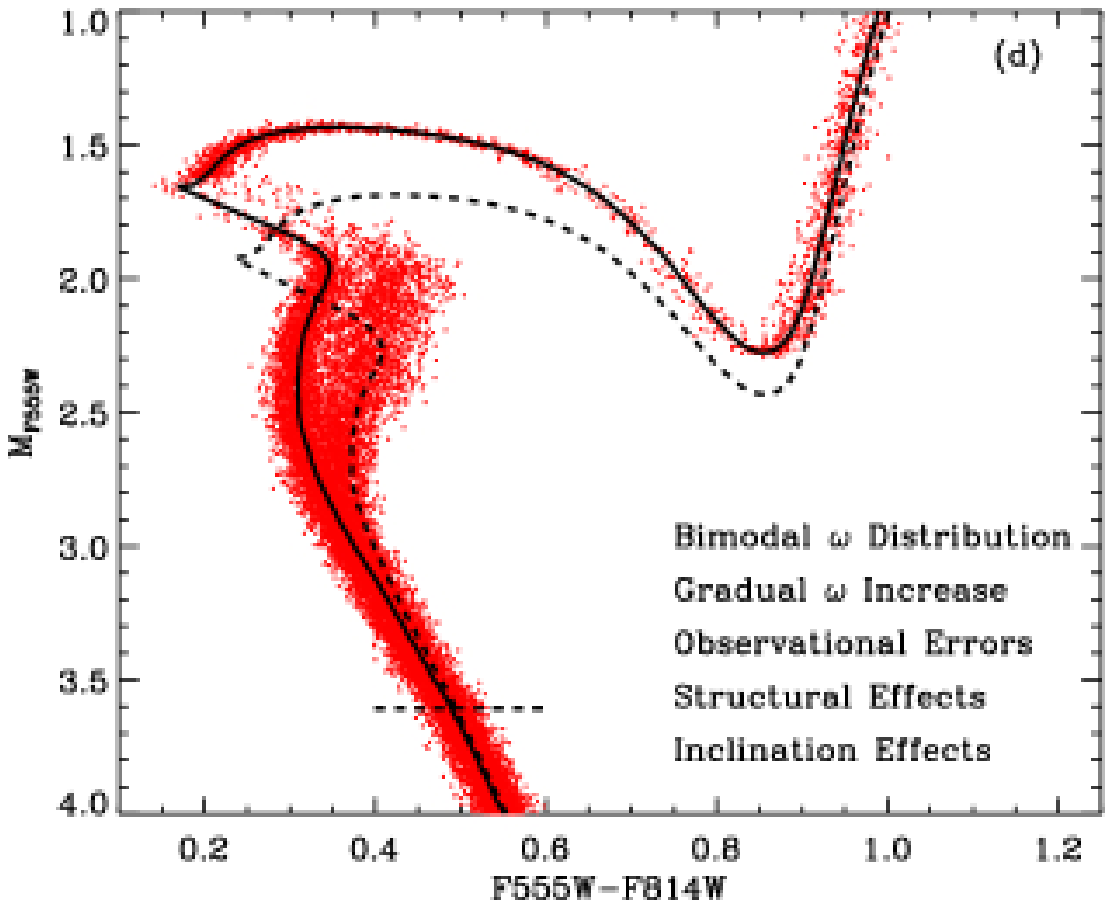}

\caption{Simulated CMDs including the effects of rotation.  The
isochrone for an age of 1.25~Gyr is shown as a solid line.  In each
panel the effects that are included in the simulation are given (see
text for details).  The horizontal dashed line marks the position of a
1.2\msun\ star.  Note that in the panel (a) all stars with masses
outside the range allowed for rotation (see text) are located behind
the isochrone.  In all panels we also show an isochrone with an age of
1.5~Gyr and metallicity of z=0.008 as a dashed line for reference.}  \label{fig:results} \end{figure*}

Panel (a) in Fig.~\ref{fig:results} shows the sampled CMD affected by
rotation as described in the previous section. The horizontal dashed
line indicates the location of a 1.2\msun\ star, below which we do not
consider rotation.  The dominant effect responsible for the spread is
the reduction of the temperature due to reduced centrifugal
force. Although the change in effective temperature is small, the
steep dependence of the V-I colour on the effective temperature
magnifies the spread in the CMD. The effect of the inclination is only
significant for those stars viewed nearly pole- or equator-on and is
always about a factor 6 smaller than the change due to structural
effects of rotation, but it is responsible for the stars located on
the blue side of the non-rotating isochrone.
In panel (b) we include typical observational errors, taken to be
normally distributed with a standard deviation of 0.01 and 0.014 in
magnitude and colour respectively. These errors are derived from the
spread in the main-sequence band of observed clusters. We note that
these errors may be overestimated as a part of this spread may be
caused by rotational broadening of the main sequence.

For clarity and simplicity we have assumed so far that an abrupt
transition occurs at 1.2\msun; lower mass stars are assumed to be
non-rotating due to magnetic braking (however, see Footnote~1).
In reality the transition is more gradual.  For example, in solar-type
stars in our own galaxy magnetic fields are detected up to
$\sim1.5$\msun (Donati \& Landstreet 2009).  Due to the large distance
to the LMC, magnetic fields and rotation rates for low mass stars are
hard to detect directly.  To show the effect of a more gradual
transition we multiply $\omega$ for each star by a factor, which
depends linearly on the mass, such that there is no rotation at
$1.2\msun$ and at $1.5\msun$ each star has the full rotation assigned
to it.  Under this assumption the spread in the CMD of an
intermediate-age cluster is limited to the turn-off region, as shown
in panel (c), which is similar to what is observed for many
intermediate-age LMC clusters (c.f.~Milone et al.~2009).  Although the
resemblance is striking, we warn the reader that the assumed
mass-dependence of the velocity distribution here is adhoc.

Royer et al.~(2007) claim that the observed distribution of rotational
velocities for A-type stars is bimodal.  Tidal torques in close
binaries and magnetic torques in Ap stars will cause some stars to be
slow rotators ($\omega <0.1$) while many stars have considerable
rotation rates ($\omega \ge 0.5$).  Therefore, we simulate a cluster
assuming a bimodal $\omega$ distribution, namely two gaussians with
equal number of stars with peaks at 0.1 and 0.6 with standard
deviations of 0.1 (again not allowing any stars to have $\omega >
0.7$).  Panel (d) shows that a bimodal $\omega$ distributions may
result in bimodal turn-off.  We emphasize that, the precise shape
depends on the assumed distribution. In particular, the assumed width
of the 'fast rotator' peak determines the spread in the CMD of the
'red sequence': a narrow 'fast rotator' peak in the $\omega$
distribution is needed to obtain a well-defined bimodal spread around
the turn-off.

\section{Discussion and conclusions}
\label{sec:discussion}

As discussed above, stars that rotate can have significantly different effective temperatures and luminosities compared to their non-rotating counterparts, even for moderate rotation rates (20-50\% of the critical break-up rotation rate).  The general effect of this is to shift stars to the red in the colour-magnitude diagram which causes a spread to occur which is detectable with high precision photometry.  

Models of cluster CMDs that include rotation can explain the spread
observed near the turn-off of intermediate age clusters in the LMC
(Mackey \& Broby Nielsen~2007; Milone et al.~2009; Goudfrooij et
al.~2009).  Additional support for this model comes from the
observations of Milone et al.~(2009).  From their Table~3, it is clear
that all clusters older than 1.5~Gyr do not show a spread in the
observed turn-off, and all but one cluster\footnote{The exception
being NGC~1975 whose CMD is not very populated, making it difficult to
ascertain whether or not there is an intrinsic spread.} below this age
have a significant spread in their turn-off.  The models developed
here are for stars with radiative envelopes, as stars with convective
envelopes are thought to have magnetic fields which will slow down any
primordial rotation.  Hence, older clusters with turn-off masses lower
than the mass where the stellar envelopes become convective, are not
expected be significantly affected by stellar rotation.

We emphasize that a large spread around the turn-off can be explained
without assuming extreme rotation rates (we assumed a gaussian
distribution in $\omega$ with a peak at 0.4 and a standard deviation
of 0.25 - with an imposed maximum of 0.70). The number of fast
rotating stars may be larger than what we have assumed.  Royer et
al.~(2007) find that the distribution of F0-F2 stars ($\sim1.4\msun$)
is bimodal with peaks of $\omega$ at 0.1 and 0.5, with the 0.5 peak
being significantly stronger (their Fig.~10).  If we would include
such a distribution in our model, the spread in the turn-off becomes even
larger.


The model that we have introduced makes specific predictions which, in
principle, can be tested.  The clearest prediction is that the stars
furthest from the main-sequence should be fast rotators.  However, as
the effect of inclination is smaller than the effect on the stellar
structure, fast-rotating stars far from the nominal main-sequence may appear
to be slow rotators if they are observed nearly pole-on, so large
samples will be required.  Stars near the turn-off of intermediate age
LMC clusters, with $V=20.5-21.5$, are near, but within the
observational limits of current 8-10m class facilities.
Alternatively, these effects should also be visible in
intermediate-age clusters in the Galaxy where the turn-off is more
readily accessible.  Measurements by Fossati et al.~(2008) for stars
in the Praesepe cluster ($\sim700$~Myr) show some evidence for rapidly
rotating stars to lie red-ward of the nominal main sequence,
consistent with the results presented here.

Additionally, fast rotating stars near the turn-off are expected to
have modified surface abundances: fragile elements such as Li, B and
Be will be depleted (Charbonnel \& Talon 1999) and in very fast
rotators N may be enhanced and C depleted.

Due to the steep dependence of (V-I) on effective temperature,
we do not expect all young clusters to show broadened turn-offs.  For
younger clusters, the main sequence turnoff occurs at higher
temperatures, where the relation between $\Delta (V-I)$ and
$\Delta(\log T_{\rm eff})$ is much shallower.  However, in these young
clusters we would predict that the main sequence (in the mass range of
$\sim1.2-1.7\msun$) would be broader than that expected from
observational errors (in the same way that the turn-off is broader in
intermediate age clusters).

We emphasize that a number of simplifications and assumptions have
been adopted in order to estimate the effect of rotation on observed
colour-magnitude diagrams. We have adopted an adhoc treatment of
magnetic braking for lower mass stars. This assumption affects the
shape of the spread below the turn-off, but it does not affect our
main conclusion that the effects of rotation are important near the
turn-off. We have also ignored that rotation can slightly prolong the
lifetime of the star, which would cause an additional spread around
the turn-off.  Furthermore, we limited this investigation to effects
of rotation on turn-off and main-sequence stars.  Just beyond the end
of the main sequence stars contract and will rotate even faster. This
evolutionary phase is fast compared to the main-sequence evolution and
not many stars are expected to be in this stage.  
Also we emphasize that
interacting binaries can complicate the picture. These effects will be
discussed in more detail in a forthcoming paper.

However, the results obtained clearly demonstrate that rotation can
significantly affect the observational properties of an intermediate age cluster, hence
detailed modeling of rotational effects and magnetic braking of stars
in the mass range considered ($1.2-1.7$\msun) would provide valuable
insight in the interpretation of observed CMDs and possible biases in
age derivations based on CMD fitting.

Finally, we note that rotation is unlikely to be the cause of the multiple stellar populations observed in old globular clusters, as low mass stars ($< 1$\msun) are not expected to be rapid rotators.


\section*{Acknowledgments}

We gratefully acknowledge John Eldridge for help in converting the
change in effective temperature into colours, Dougal Mackey and
Antonino Milone for sending us their data and Georges Meynet for
providing unpublished rotating stellar models.
Furthermore we are grateful to Onno Pols, Norbert Langer, Rob Rutten,
Ines Brott, Matteo Cantiello, Eugenio Carretta, Angela Bragaglia, Chris Evans and
the referee Corinne Charbonnel for their comments.


\bsp
\label{lastpage}

\begin{thebibliography}{99}
\bibitem[Bekki  \& Mackey(2009)]{2009MNRAS.394..124B} Bekki, K., \& Mackey, A.~D.\ 2009, MNRAS, 394, 124
\bibitem[Boesgaard(1987)]{1987PASP...99.1067B} Boesgaard, A.~M.\ 1987, PASP, 99, 1067 
\bibitem[Cardini \& Cassatella(2007)]{2007ApJ...666..393C} Cardini,
D., \& Cassatella, A.\ 2007, ApJ, 666, 393
\bibitem[Charbonnel \& Talon(1999)]{1999A&A...351..635C} Charbonnel, C., \& Talon, S.\
1999, A\&A, 351, 635

\bibitem[Collins \& Smith(1985)]{1985MNRAS.213..519C} Collins, G.~W.,
II, \& Smith, R.~C.\ 1985, MNRAS, 213, 519
\bibitem[De Mink et al.(2009)]{2009A&A...497..243D} De Mink, S.~E., Cantiello, M., Langer, N., Pols, O.~R., Brott, I., \& Yoon, S.-C.\ 2009, A\&A, 497, 243
\bibitem[D'Ercole et al.(2008)]{2008MNRAS.391..825D} D'Ercole, A., 
Vesperini, E., D'Antona, F., McMillan, S.~L.~W., \& Recchi, S.\ 2008, MNRAS, 391, 825 
\bibitem[Donati \& Landstreet(2009)]{2009arXiv0904.1938D} Donati, J., \& Landstreet, J.\ 2009, ARAA (arXiv:0904.1938)
\bibitem[D'Souza et al.(1992)]{1992JApA...13..109D} D'Souza, J., Mathew, 
A., \& Rajamohan, R.\ 1992, Journal of Astrophysics and Astronomy, 13, 109 

\bibitem[Eldridge et al.(2007)]{2007MNRAS.376L..52E} Eldridge, J.~J., Mattila, S., \& Smartt, S.~J.\ 2007, MNRAS, 376, L52 
\bibitem[Faulkner et al.(1968)]{1968ApJ...151..203F} Faulkner, J., 
Roxburgh, I.~W., \& Strittmatter, P.~A.\ 1968, ApJ, 151, 203 

\bibitem[Fossati et al.(2008)]{2008A&A...483..891F} Fossati, L., Bagnulo, S., Landstreet, J., Wade, G., Kochukhov, O., Monier, R., Weiss, W., \& Gebran, M.\ 2008, A\&A, 483, 891 
\bibitem[Gaige(1993)]{1993A&A...269..267G} Gaige, Y.\ 1993, A\&A, 269, 267 

\bibitem[Glatt et al.(2008)]{2008AJ....136.1703G} Glatt, K., et al.\ 2008, AJ, 136, 1703 

\bibitem[Goudfrooij et al.(2009)]{2009AJ....137.4988G} Goudfrooij, P., 
Puzia, T.~H., Kozhurina-Platais, V., \& Chandar, R.\ 2009, AJ, 137, 4988
 
\bibitem[Gray  \& Garrison(1987)]{1987ApJS...65..581G} Gray, R.~O., \& Garrison, R.~F.\ 1987, ApJS, 65, 581 

\bibitem[Heger et al.(2000)]{2000ApJ...528..368H} Heger, A., Langer, N., \& Woosley, S.~E.\ 2000, ApJ, 528, 368


\bibitem[Huang \& Gies(2006)]{2006ApJ...648..591H} Huang, W., \& Gies,
D.~R.\ 2006, ApJ, 648, 591

\bibitem[Huang \& Gies(2008)]{2008ApJ...683.1045H} Huang, W., \& Gies,
D.~R.\ 2008, ApJ, 683, 1045

\bibitem[Hunter et al.(2008)]{2008A&A...479..541H} Hunter, I., Lennon,
D.~J., Dufton, P.~L., Trundle, C., Sim{\'o}n-D{\'{\i}}az, S., Smartt,
S.~J., Ryans, R.~S.~I., \& Evans, C.~J.\ 2008, A\&A, 479, 541

\bibitem[Mackey \& Broby Nielsen(2007)]{2007MNRAS.379..151M} Mackey, A.~D., \& Broby Nielsen, P.\ 2007, MNRAS, 379, 151 
\bibitem[Mackey et al.(2008)]{2008ApJ...681L..17M} Mackey, A.~D., Broby 
Nielsen, P., Ferguson, A.~M.~N., \& Richardson, J.~C.\ 2008, ApJL, 681, L17 
\bibitem[Massey \& Hunter(1998)]{1998ApJ...493..180M} Massey, P., \& Hunter, D.~A.\ 1998, ApJ, 493, 180 
\bibitem[Mestel \& Spruit(1987)]{1987MNRAS.226...57M} Mestel, L., \& Spruit, H.~C.\ 1987, MNRAS, 226, 57 
\bibitem[Meynet \& Maeder(1997)]{1997A&A...321..465M} Meynet, G., \& Maeder, A.\ 1997, A\&A, 321, 465

\bibitem[Milone et al.(2009)]{2009A&A...497..755M} Milone, A.~P., Bedin, L.~R., Piotto, G., \& Anderson, J.\ 2009, A\&A, 497, 755 

\bibitem[Palacios et al.(2003)]{2003A&A...399..603P} Palacios, A., Talon, S., Charbonnel, C., \& Forestini, M.\ 2003, A\&A, 399, 603


\bibitem[P{\'e}rez Hern{\'a}ndez et al.(1999)]{1999A&A...346..586P}
P{\'e}rez Hern{\'a}ndez, F., Claret, A., Hern{\'a}ndez, M.~M., \&
Michel, E.\ 1999, A\&A, 346, 586

\bibitem[Pietrinferni et al.(2004)]{2004ApJ...612..168P} Pietrinferni, A., Cassisi, S., Salaris, M., \& Castelli, F.\ 2004, ApJ, 612, 168 
\bibitem[Piotto(2008)]{2008MmSAI..79..334P} Piotto, G.\ 2008, Memorie della 
Societa Astronomica Italiana, 79, 334 (arXiv:0801.3175)
\bibitem[Royer et  al.(2007)]{2007A&A...463..671R} Royer, F., Zorec, J., \& G{\'o}mez, A.~E.\ 2007, A\&A, 463, 671 
\bibitem[Salpeter(1955)]{1955ApJ...121..161S} Salpeter, E.~E.\ 1955, ApJ, 121, 161 
\bibitem[Schatzman(1962)]{1962AnAp...25...18S} Schatzman, E.\ 1962, Annales 
d'Astrophysique, 25, 18 
\bibitem[Talon  \& Charbonnel(2004)]{2004A&A...418.1051T} Talon, S., \& Charbonnel, C.\ 2004, A\&A, 418, 1051 
\bibitem[Townsend et al.(2004)]{2004MNRAS.350..189T} Townsend, R.~H.~D., 
Owocki, S.~P., \& Howarth, I.~D.\ 2004, MNRAS, 350, 189 

\bibitem[Westera et al.(1999)]{1999ASPC..192..203W} Westera, P., Lejeune, 
T., \& Buser, R.\ 1999, Spectrophotometric Dating of Stars and Galaxies, ASP conference series 192, 203 

\bibitem[von Zeipel(1924)]{1924MNRAS..84..665V} von Zeipel, H.\ 1924, MNRAS, 84, 665 
\bibitem[Yoon et al.(2006)]{2006A&A...460..199Y} Yoon, S.-C., Langer, N., \& Norman, C.\ 2006, A\&A, 460, 199 












\end{thebibliography}
\end{document}